**Adhesion Mediated by Competition of Ligand-Receptor Binding Against the Lateral Osmotic Pressure of Mobile Repellers**

A. Boulbitch

Dept. for Physics E22, technical University Munich, Gemany

**Abstract**

A model system has been recently developed to study adhesion. It consists of a giant lipid bilayer vesicle with reconstituted lipo-polymers (repellers) as well as with lipo-ligands recognized by receptors covering the substrate[1]. Adhesion in this system is studied theoretically. The state of the weak adhesion is shown to be dominated by the competition of the gravitation and the undulation repulsion between the membrane and the substrate. The state of the tight adhesion is formed by a competition of the ligand-receptor binding and the lateral osmotic pressure of the mobile repeller molecules. The regions of the weak and the tight adhesion are separated on the phase diagram by a whole transition region due to the scattering in parameters of vesicles.

## 1. Introduction

Though adhesion of cells represents an essential stage in various biological processes[2] its physical properties are still poorly understood. Physics basis of the membrane adhesion mediated by generic interactions has been theoretically studied in the papers[3-6]. Phase separation induced by the non-specific adhesion has been described in the papers[7, 8].

Since the theory of Bell[9] it is recognized that adhesion of cells is due to the ligand-receptor binding, rather than due to generic forces. Nevertheless, only few theoretical works have been undertaken to study specific adhesion of biomembranes. It has been shown that membranes adhere, if the concentration of the adhesive stickers exceeds a certain threshold[10]. Unbinding by the way of rupture of multiple parallel bonds has been studied relating the rupture force and the number of bonds[11, 12]. It has been shown that adhesion at low receptor densities may take place by first order adhesion transition accompanied by receptor segregation[13]. The generic repulsion between the membrane and the substrate has been shown to give rise to a lateral phase separation of adhesive stickers[14]. Nucleation during the specific adhesion has been studied in the paper[15]. Kinetics of the specific adhesion has been analyzed in the papes[16-19].

Already in the pioneering paper[20] it has been shown that in a simple system consisting of two membranes carrying ligands and receptors the adhesion is controlled by the lateral osmotic pressure of ligands, receptors and ligand-receptor pairs[19]. In addition to ligands, receptors and pairs membranes of adhering cells contain however, a number of different species which are neither ligands nor receptors such as various membrane proteins and sugars forming the glycocalix and penetrating into the extracellular space. Some of them are immobilized via anchoring to the cytoskeleton, while others possess a lateral mobility[2, 21]. One mechanism of the influence of these species on adhesion is a short-range steric repulsion of the cell membranes $\sim \exp(-z/D)/z$ which effect on adhesion has been discussed in the paper[20], where $z$ is the inter-membrane distance and $D$ is the length of the glycocalix. Such a repulsion arises between the approaching cells, as soon as their glycocalices with no lateral mobility overlap. In this paper we discuss another mechanism impeding the adhesion and related to the laterally mobile non-reacting species.

Adhesion *in-vivo* takes place by overlapping of a number of phenomena which makes it difficult to study its physical basics. To understand these phenomena it is essential to establish simplified experimental models which would contain main components of cell membranes and "catch" main feathers of the bioadhesion. Recently such an experimental model system has been developed[1, 17, 22-26]. This system (referred to as the **"model system"**) is studied theoretically in the present paper. The model system is shown schematically in Fig. 1.

It consists of a giant lipid bilayer vesicle (Fig. 1 i) which acts as a test cell. The vesicle carries the lipid-coupled-ligands (the cyclic RGD motif has been used in[1, 17, 22-26]) shown in Fig. 1 (jj). In addition, lipo-polymers (such as the polyethyleneglycol covalently attached to the lipid head-group) have been reconstituted into the vesicle (Fig. 1 j). The latter has been introduced in order to mimic the repulsive effect of the glycocalix[1, 17, 22-26] and for this reason is referred to as "repellers". We show below however, that the main effect of the repellers is not the repulsion. To mimic the target cell a substrate coated by the immobilized receptors (the proteins $\alpha_{IIb}\beta_3$-integrins[1]) recognizing the ligands was used (Fig. 1 jjj).

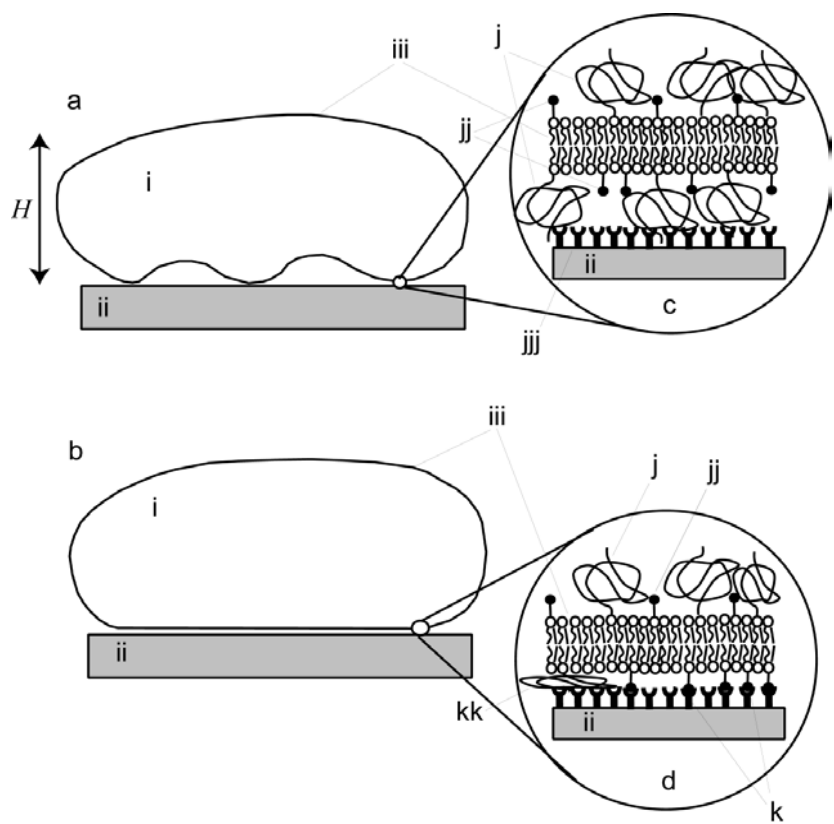

Fig. 1. Schematic view of the experimental model system consisting of a giant lipid bilayer vesicle with the reconstituted ligands and repeller molecules and a substrate coated by the receptors. The system exhibits the weak (a) and the tight (b) states of adhesion. (c) and (d) display the increased view of the membrane and the substrate in the states (a) and (b). (i) shows the vesicle, (ii) the substrate, (iii) the lipid bilayer, (j) the lipo-polymers (repellers), (jj) the lipo-ligands, (jjj) receptors, (k) ligand-receptor pairs, (kk) the squeezed mushroom trapped inside the area of the tight adhesion.

With this new system the specific statics[1, 22-26] and kinetics[17] of the adhesion has been studied. Two types of the adhesion states referred to as the "weak" and the "tight" adhesion with a discontinuous transition between them have been observed[1]. The state of the weak adhesion exhibits manifested fluctuations of the membrane adjacent to the substrate. The membrane-substrate distance in the weak state has been reported in the range of ~100nm [17]. In contrast, in the state of the tight adhesion the membrane-substrate distance is equal to the size of the head-group of the ligands and no fluctuations of the membrane adjacent to the substrate have been observed[1, 17]. Variation of the repeller concentration enabled one to switch between the tight and the weak adhesion states. On the plane of the repeller concentration versus the concentration of ligands the region with a high concentration of repellers and small concentration of ligands corresponds to the state of the weak adhesion. In the opposite corner of the phase diagram the tight adhesion state takes place. These two regions are separated by a relatively wide region in which the weak adhesion co-exists with the tight one. The value of the adhesion energy has been measured. It appeared to be close to the osmotic pressure of the repellers[1].

In the present paper we address the above phenomena. We give a theoretical description of the unbinding phase diagram of the model system and discuss the extraordinary role of the laterally mobile repeller molecules.

The paper is organized as follows. In the Section 2 the state of the weak adhesion is described. In the Section 3 the free energy of the tight adhesion is discussed. In the Section 4 the free energy is analyzed and the phase diagram is built. Section 5 contains estimates and in Section 6 we discuss the results.

## 2. The state of the weak adhesion

As reported in the paper[1], depending on the concentration of ligands and repellers, vesicles adhered on the substrate can be in one of two equilibrium states of adhesion. The first of these states is characterized by a relatively large (~100nm) distance between the membrane and the substrate and manifested membrane fluctuations (Fig. 1 a).

In the beginning of the experiment the vesicles are hovering over the substrate and sink towards the substrate due to the slight difference $\Delta\rho$ in the density of the of the fluid inside with respect to that outside the vesicle. The repeller molecules form mushrooms the Flory radius of which, $R_F$, is 3.5nm. This is enough to screen down non-specific membrane-substrate interactions (such as electrostatics and van der Waals forces). In addition $R_F$ is larger than the length of the head-group of the ligand (2.2nm)[17]. For this reason the repellers prevent also the ligand-receptor binding. The only forces acting in the system are the Helfrich repulsion[27] and the gravitation yielding the membrane energy:

$$U \sim \frac{(k_B T)^2}{\kappa z^2} + \Delta\rho g H z \qquad (1)$$

where $T$ is the temperature, $\kappa$ is the membrane bending rigidity, $z$ is the distance between the membrane and the substrate and $g \approx 9.8$ m/s$^2$. The vesicles in experiments reported in the papers[1, 17] formed ellipsoidal caps on the substrate with the semi-axis ratio close to 2. In this case the vesicle height, $H$, varies relatively slowly with position within the adhesion interface and for the purposes of the estimate can be approximately considered as a constant. Minimization of the potential Eq. (1) yields the equilibrium height of the membrane over the substrate:

$$z_{\min} \sim \left[ \frac{(k_B T)^2}{\kappa g H \Delta\rho} \right]^{1/3} \qquad (2)$$

**3. Free energy of the membrane adhesion mediated by ligand-receptor binding**

In the tight adhesion state the membrane-substrate distance is determined by the length of the ligand head-group. The latter is smaller than the sizes of the repeller mushrooms. The mushrooms therefore, must be pushed out from the adhesion area. The free energy of the membrane in the state of the tight adhesion can be written as the sum of four terms:

$$F_{tight} = F_L + F_R + F_P + F_{REP} \quad (3)$$

where $F_L$, $F_R$, $F_P$ and $F_{REP}$ are the free contributions of ligands, receptors, the ligand-receptor pairs and of the repeller molecules.

### 3.1. Free energy of ligands, receptors and ligand-receptor pairs

The first three terms of the free energy (3) have the form:

$$F_L = \mu_L (N_L - N_P) + k_B T (N_L - N_P) \ln\left[\frac{N_L - N_P}{eN}\right] \quad (4)$$

$$F_R = \mu_R (N_R - N_P) \quad (5)$$

$$F_P = \mu_P N_P + k_B T N_P \ln\left[\frac{N_P}{eN}\right] \quad (6)$$

where $\mu_L$, $\mu_R$ and $\mu_P$ are the specific chemical potentials of the ligands, receptors and ligand-receptor pairs correspondingly, $N_L$, $N_R$ and $N_P$ are the initial numbers of the ligands and receptors and the current number of the ligand-receptor pairs. $N$ is the number of the solvent molecules (lipids). In the present contribution we give a description of the phenomena reported in the paper [1] in which the receptor molecules were immobilized on the substrate by physisorbtion. Accordingly, the receptor free energy (5) contains no entropic terms. Equations (4)-(6) describe the adhesion in the spirit of the approach of the paper[20].

### 3.2. The contribution of the repellers to the free energy.

Since the mushrooms formed by the repeller molecules have no place in the adhered area, the fraction γ of them is pushed out of the region where the membrane is tightly adhered to the substrate and the membrane-substrate distance is equal to the length of the RGD head-group. The work of the squeezing the repellers out of the area of the tight adhesion has the form:

$$F_{\text{REP}} = -\gamma \int_A^{A-A_*} \Pi dA = -\gamma \Pi_0 A \ln(1 - A_*/A) \tag{7}$$

where $\Pi$ is the lateral osmotic pressure of the repellers and the integration runs from the total vesicle surface area $A$ which is accessible for the repellers in the state of the weak adhesion to the area $A - A_*$ which the repellers can occupy in the tight adhesion state. Here $A_*$ is the area of the tightly adhered part of the vesicle membrane. The value $\Pi_0$ of the lateral osmotic pressure corresponds to the weakly adhered state of the vesicle (i.e. if $A_* = 0$):

$$\Pi_0 = \rho_{\text{REP}} k_B T \tag{8}$$

where $\rho_{\text{REP}} = N_{\text{REP}}/A$ is the surface number density of the repellers, $N_{\text{REP}}$ is their number and the factor $\gamma$ accounts for probability for some repellers to stay inside the adhesion area in a deformed state (as it is schematically shown in Fig. 1 b (kk)):

$$\gamma = 1 - \exp(-\Delta E / k_B T) \tag{9}$$

Here $\Delta E$ is the energy of deformation of the mushroom. It can be calculated as the energy of deformation of the Gaussian coil[28]:

$$\Delta E = \frac{3 k_B T (\Delta R)^2}{2 R_F^2} \tag{10}$$

Here $\Delta R \approx R_F - d$, where $d$ is the size of the ligand. The detailed proof is given in the Appendix.

In the following we will assume that $A_*/A \ll 1$ which yields $F_{\text{REP}} \approx \gamma \Pi_0 A_*$. If the ligand-receptor pairs are densely packed, one finds $A_* \approx a N_P$, where $a$ is the area per ligand-receptor pair, and the work function (7) can be approximated as

$$F_{\text{REP}} \approx \gamma \Pi_0 a N_P \tag{11}$$

8## 4. Phase diagram

### 4.1. Equation of state

Introduce the dimensionless concentrations of the ligands, $x_L = N_L/N$, receptors, $x_R = N_R/N$, and ligand-receptor pairs, $x_P = N_P/N$. In the regime of the tight adhesion the free energy per lipid, $f_{tight} = F_{tight}/N$, takes the form

$$f_{tight} = \mu_L x_L + \mu_R x_R + (\gamma \Pi_0 a - \varepsilon) x_P + k_B T \{x_P \ln(x_P/e) + (x_L - x_P) \ln[(x_L - x_P)/e]\} \quad (12)$$

where $\varepsilon = \mu_L + \mu_R - \mu_P > 0$. Minimization of the free energy (12) yields equation of state

$$\gamma \Pi_0 a - \varepsilon + k_B T \ln \frac{x_P}{(x_L - x_P)} = 0 \quad (13)$$

Equation (13) always has a solution which determines the equilibrium concentration $x_P$ of the ligand-receptor pairs:

$$x_P = \frac{x_L}{1 + \exp(\gamma \rho_{REP} a - \varepsilon / k_B T)} \quad (14)$$

In the state of the weak adhesion the free energy of the membrane has the form $F_{weak} = \mu_L N_L + \mu_R N_R + k_B T N_L \ln(N_L/Ne)$ yielding the free per lipid molecule

$$f_{weak} = F_{weak}/N = \mu_L x_L + \mu_R x_R + k_B T x_L \ln(x_L/e) \quad (15)$$

where we omit the energy (1) which is small with respect to the other terms. The difference $\Delta F = F_{tight} - F_{weak}$ is the free energy gain due to the tight adhesion (while $\Delta f = \Delta F/N$ is that per lipid molecule). It is related to the so-called, adhesion energy $W$ defined as the free energy gain per unit area $W = -\Delta F / A_*$. Making use of the equation of state (13) one finds

$$\Delta f = k_B T x_L \ln(1 - x_P / x_L) \quad (16)$$

Since the number of the ligand-receptor pairs is always smaller than the initial number of lipids ($x_P / x_L \leq 1$), the energy gain Eq. (16) is always negative. It is reasonable to assume that at the unbinding transition the number of the pairs is much smaller than that of ligands $x_P / x_L \ll 1$. In this case

$$\Delta f \approx -k_B T x_P \qquad (17)$$

**4.2. The unbinding transition**

The unbinding condition takes place as soon as the bending elastic energy of the membrane overcomes the adhesion energy $W$[4]. Thus the transition condition takes the form

$$\Delta F = -\kappa \omega g \qquad (18)$$

where $\kappa$ is the membrane bending modulus, $g = 4\pi A_* / A$ and following the paper[4] we introduce the dimensionless parameter $\omega = WA / 4\pi\kappa$. Both $g$ and $\omega$ depend on the parameters of the vesicle in solution (such as the osmotic pressure difference, excess area, spontaneous curvature and so on), but are independent of the concentrations of ligands and repellers. In the papers[3,4] the unbinding phase diagrams have been built in terms of the reduced adhesion energy $\omega$ *versus* the osmotic pressure difference[4] or *versus* the enclosed volume[3]. On these planes the transitions between different configurations of the vesicle in adhered and non-adhered states take place along the lines described in[3,4]. In the following we assume that the values of $\omega$ and $g$ Eq. (18) correspond to such a transition line. Making use of Eq. (17) one finds the number of the ligand-receptor pairs at the unbinding transition in terms of the parameters $\omega$ and $g$:

$$N_P^{(\text{trans})} \approx \kappa g \omega / k_B T \qquad (19)$$

Substituting this value into equation of state (14) one finds the relation between the concentration of ligands and repellers at the unbinding transition:

$$x_{\text{REP}} = \frac{a_{\text{lip}}}{\gamma a} \left( \tilde{\varepsilon} / k_B T + \ln x_L \right) \qquad (20)$$

where $a_{\text{lip}}$ is the membrane area per lipid molecule and

$$\tilde{\varepsilon} = \varepsilon - k_B T \ln x_P^{(trans)} \qquad (21)$$

Equation (20) describes the line on the phase diagram on the plane with coordinates $x_L$ and $x_{REP}$ which separates the upper part where the weak adhesion mediated by gravitation takes place from the lower part where the adhesion is due to the ligand-receptor binding (Fig. 2).

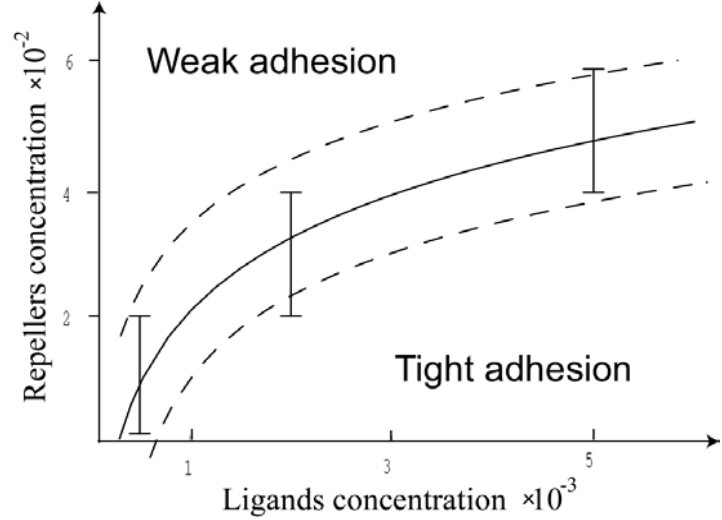

Fig. 2. Phase diagram on the plane $(x_L, x_{REP})$ describing the unbinding transition of a vesicle shown in Fig.1. The bars show the experimental data taken from the paper[1]. The solid line shows the best fit of these data making use of the equation (20). The dashed lines schematically indicate the transition region.

### 5. Estimates

Consider first the state of the weak adhesion. We have seen that this state is determined by the competition of the gravitation (pressing the membrane towards the substrate) and the Helfrich repulsion force. Existence of the Helfrich repulsion implies that this state exhibits manifested membrane fluctuations. This corresponds to the observations reported in the papers[1, 17]. Using the values of the bending elasticity $\kappa \approx 4 \times 10^{-19}\,\text{J}$ [29], the vesicle height (Fig. 1 a) $H \sim 10\,\mu\text{m}$ and the excess density of the fluid $\Delta\rho \approx 50\,\text{kg/m}^3$ from the measurements[1] one estimates the distance between the membrane and the substrate Eq. (2) in the state of the weak adhesion $z_{min} \sim 100\,\text{nm}$ which agrees with the observations (see Fig. 1 b from the paper[17]). The membrane energy in the minimum of the potential (1) is $U \sim 10^{-9} \text{J/m}^2$ which is much smaller than the adhesion energy $(\sim 10^{-5} \text{J/m}^2)^1$ in the state of the tight adhesion

Fitting the experimental phase diagram reported in the paper[1] by equation (20) yields the equation $x_{REP} \approx 0.14 + 0.02 \times \ln x_L$. Estimating the mushroom size $R_F \approx 3.5 \text{nm}$ and the length of the ligand $d \approx 2.2 \text{nm}$ [17] one finds the estimate $\gamma \approx 0.19$. Estimating $a_{lip} \approx 0.6 \text{nm}^2$ and $a \approx 240 \text{nm}^2$ [30] one finds the ratio $a_{lip}/\gamma a \approx 0.01$ in good correspondence with the value 0.02 obtained by the fitting.

Thus the estimates show a good agreement between the theoretical results and the measurements[1, 17].

## 6. Discussion

The thermodynamics of the adhesion mediated by ligand-receptor binding has been considered for the first time in the paper [20]. In this paper the continuos unbinding transition controlled by a repulsion between the cell surfaces has been described. The repulsion potential $\sim \exp(-z/D)/z$ has been attributed to the combination of the osmotic pressure of the solvent being squeezed out from the space between the two cells and the steric compression of the glycocalix[20]. The effect of the membrane bending on the unbinding in the paper[20] has been neglected.

The unbinding transition mediated by the competition of the membrane adhesion and bending has been studied within the example of the unbinding of vesicles adhered via generic forces in the papers of Lipowsky and Seifert [3-5]. It has been shown that in this case depending on the parameters of the adhered vesicle the transition can be discontinuous as well as continuos.

In the model system studied here the membrane bending is important as it has been shown in the paper[1]. For this reason the results[3-5] apply also to this system. In contrast to the adhesion by generic forces[3-5], adhesion in the model system is due to the ligand-receptor binding like in the one studied by Bell and co-authors[20]. The difference between the model of Bell[20] and the one studied here is due to different behavior of the repellers in these two models. The repeller molecules in the model system are mobile while in the model[20] the glycocalix is motionless. During the process of ligand-receptor binding the mobile repellers are mainly pushed out from the area of adhesion, rather than stay inside the adhesion area in a deformed state. The surface density of the repellers outside of the adhesion area is thus, higher than that inside and they exert a lateral osmotic pressure on the adhesion rim which tends to decrease the adhesion area. As soon as this pressure overcomes the free energy gain due to the ligand-receptor binding, the vesicle unbinds (i.e. comes into the state of the weak adhesion). This condition determines the line Eq.(20) on the phase diagram on which the transition takes place. In the vicinity of this line the value of the adhesion energy is close to that of the osmotic pressure of the repellers. This explains the observation reported in the paper[1]. The shape of the transition line (i.e. the linear relation between $x_{REP}$ and $\ln x_L$) is universal for all vesicles. It is the consequence of the fact that repellers and ligands represent two-dimensional gases. The position of this line however, depends on the parameters of the vesicle. The adhesion energy $W$ enters the expression for the transition line through $\tilde{\varepsilon}$ (19) and (21) i.e. through the chemical potential of the ligand-receptor pairs.

The first term in the right-hand-part of Eq. (20) depends on the parameters characterizing the vesicle state (such as the vesicle total surface area, osmotic pressure difference, spontaneous curvature of the membrane and so on) through the dependence of $\tilde{\varepsilon}$ on $x_P^{(trans)}$ and thus, on $g$ and $\omega$. In present there no way is known to control these parameters independently. For this reason even for vesicles obtained during the same preparation $\tilde{\varepsilon}$ exhibits a dispersion. In contrast, the second term in the right-hand-part depends only on the well-controlled parameters. Therefore, different vesicles exhibit the unbinding transition on different lines with the same shapes. Thus, instead of the sharp (continuos or discontinuous) transition line a whole transition region arises on the phase diagram. Within this region the vesicles with the same values of $x_{REP}$ and $x_L$ may be either bound or unbound depending on their mechanical parameters. The considerable transition region (rather than the transition line) has been reported in the paper[1] and shown in Fig. 2 by the bars. In the paper[1] the existence of this region led to a conclusion that the transition is of a first order type. One can see however, that due to the scattering in the value $\tilde{\varepsilon}$ the transition region will be observed also in the case of the continuos unbinding.

## 7. Summary

We studied theoretically the model system consisting of a giant vesicle with reconstituted ligands and repellers adhering on the substrate covered by receptors. The state of weak adhesion of this system is mediated by the competition of the gravitation and undulation forces. The state of the tight adhesion forms by competition of binding of the ligands and receptors on one hand and the lateral osmotic pressure of repellers on the other hand. The plane $(x_L, x_{REP})$ can be separated into three regions: (i) the stability region of the state of the weak adhesion, (ii) the region where the tight adhesion takes place and (iii) the transition region in which the vesicle may be either in the adhered or in the unbound state depending on parameters.

**Acknowledgment:** this work was supported by the grant of the German Research Society (Deutsche Forschungsgemeinschaft) Nr. SA246/28-4

## Appendix

Here we give a proof of validity of Eq.(8), (9). Consider the equilibrium state of the repellers dissolved in the two-dimensional lipid liquid. The key point is that the adhesion rim 14divides the whole membrane into two parts (the adhesion and the free regions) and represents a semi-permeable boundary both for lipids and repellers. Denote the parameters belonging to the adhesion region by the superscribe asterix "$*$", while those of the free membrane part will carry the superscribe "f". The chemical potential of the solvent (lipids) can be written as[31]

$$\mu_{\text{lip}} = \mu_{\text{lip}}^{(0)} - k_B T x_{\text{REP}} \qquad (22)$$

while that of the solute (repellers) has the form

$$\mu_{\text{REP}} = \mu_{\text{REP}}^{(0)} + k_B T \ln x_{\text{REP}} \qquad (23)$$

Since there is the exchange both by the solute and the solvent across the adhesion rim, equations of equilibrium take the form

$$\mu_{\text{lip}}^{(0)*} - \mu_{\text{lip}}^{(0)\,f} = k_B T \left( x_{\text{REP}}^* - x_{\text{REP}}^f \right); \quad \mu_{\text{REP}}^{(0)*} - \mu_{\text{REP}}^{(0)\,f} = k_B T \ln \left( x_{\text{REP}}^f / x_{\text{REP}}^* \right) \qquad (24)$$

From the second Eq.(24) one finds $x_{\text{REP}}^f / x_{\text{REP}}^* = \exp\{(\mu_{\text{REP}}^{(0)*} - \mu_{\text{REP}}^{(0)\,f})/k_B T\}$. The difference in the chemical potentials of the repellers inside and outside of the adhesion region is equal to the energy of the deformation of the mushroom Eq.(10): $\mu_{\text{REP}}^{(0)*} - \mu_{\text{REP}}^{(0)\,f} = \Delta E$. The variation of the chemical potentials of the solvent inside and outside the adhesion zone can be represented as $\mu_{\text{lip}}^{(0)\,f} - \mu_{\text{lip}}^{(0)*} = \left( \partial \mu_{\text{lip}}^{(0)} / \partial \Pi \right)_T \Delta \Pi$. Since $\left( \partial \mu_{\text{lip}}^{(0)} / \partial \Pi \right)_T = a_{\text{lip}}$ is the area per lipid molecule, one finds

$$\Delta \Pi = k_B T x_{\text{REP}} \left\{ 1 - \exp\left( -\Delta E / k_B T \right) \right\} a_{\text{lip}}^{-1} \qquad (25)$$

It is easy to see that $x_{\text{REP}} / a_{\text{lip}} = N^f / N a_{\text{lip}}$ and $N a_{\text{lip}} \approx A$. Thus $x_{\text{REP}} / a_{\text{lip}} \approx \rho_{\text{REP}}$ and Eq.(25) yields Eq.(8), (9).